\documentclass[twocolumn,useAMS,usenatbib]{mn2e}

\usepackage{amssymb}
\usepackage{amsmath}
\usepackage{graphicx}
\usepackage{textcomp}
\usepackage{natbib}
\usepackage{booktabs}
\usepackage{tabularx}
\usepackage{subfigure}
\usepackage{color}

\title[]{Forecasts on neutrino mass constraints from the redshift-space two-point correlation function}
\author[F. Petracca, et al.]{F. Petracca$^{1,2,3}$\thanks{E-mail:fernanda.petracca2@unibo.it},
F. Marulli$^{1,2,3}$,
L. Moscardini$^{1,2,3}$,
A. Cimatti$^{1,2}$,
C. Carbone$^{4,3}$,
\newauthor
and R. E. Angulo$^{6}$\\
$^{1}$Dipartimento di Fisica e Astronomia, Alma Mater Studiorum-Universit\`a di Bologna, viale Berti Pichat 6/2, I-40127 Bologna, Italy\\
$^{2}$INAF, Osservatorio Astronomico di Bologna, via Ranzani 1, I-40127 Bologna, Italy\\
$^{3}$INFN, Sezione di Bologna, viale Berti Pichat 6/2, I-40127 Bologna, Italy\\
$^{4}$INAF, Osservatorio Astronomico di Brera, via Bianchi 46, I-23807
Merate (LC), Italy\\
$^{5}$Centro de Estudios de F\'isica del Cosmos de Arag\'on, Plaza San Juan 1, Planta-2, E-44001 Teruel, Spain \\
}
\begin{document}

\date{Accepted -- --. Received -- --; in original form -- --}

\pagerange{\pageref{firstpage}--\pageref{lastpage}} \pubyear{2015}

\maketitle

\label{firstpage}
\begin{abstract}
We provide constraints on the accuracy with which the neutrino mass
fraction, $f_{\nu}$, can be estimated when exploiting measurements of
redshift-space distortions, describing in particular how the error on
neutrino mass depends on three fundamental parameters of a
characteristic galaxy redshift survey: density, halo bias and volume.
In doing this, we make use of a series of dark matter halo catalogues
extracted from the {\small BASICC} simulation.  The mock data are
analysed via a Markov Chain Monte Carlo likelihood analysis.  We find
a fitting function that well describes the dependence of the error on
bias, density and volume, showing a decrease in the error as the bias
and volume increase, and a decrease with density down to an almost
constant value for high density values.  This fitting formula allows
us to produce forecasts on the precision achievable with future surveys
on measurements of the neutrino mass fraction.  For example, a
Euclid-like spectroscopic survey should be able to measure the
neutrino mass fraction with an accuracy of $\delta f_{\nu} \approx
6.7\times10^{-4}$, using redshift-space clustering once all the other
cosmological parameters are kept fixed to the $\Lambda$CDM case.
\end{abstract}

\begin{keywords}
neutrino relics -- cosmological parameters -- dark energy -- large-scale structure of the Universe.
\end{keywords}

\section{Introduction}
Estimating the neutrino mass is one of the main challenges of
cosmology today.  According to the standard model of particle physics,
neutrinos are weakly interacting massless particles.  However, the
experiments on the oscillations of solar and atmospheric neutrinos
tell us that neutrinos cannot be massless.  Oscillation experiments
can only measure the differences in the squared masses of the neutrino
eigenstates $(m_1, m_2, m_3)$ and not the absolute mass scale. The
current data imply $|\Delta m_{31}^2|\simeq 2.4 \times
10^{-3}\ \mathrm{eV}^2$ and $\Delta m_{21}^2\simeq 27.6 \times
10^{-5}\ \mathrm{eV}^2$ \citep{2012PhRvD..86a0001B}.  These measurements
provide a lower limit for the sum of neutrino masses of $\approx
0.06\ \mathrm{eV}$ (see \citealt{2014NJPh...16f5002L} for a review).

Now that cosmology has entered the ``precision era'' and the
cosmological parameters can be constrained at a percent level,
observations of the Universe can assist in the quest for neutrino
mass, since neutrinos affect the evolution of the universe in several
observable ways.

After thermal decoupling, relic neutrinos constitute a collisionless
fluid, where the individual particles free-stream with the
characteristic thermal velocity.  As long as neutrinos are
relativistic, the free-streaming scale is simply the Hubble radius.
When they become non-relativistic, their thermal velocity decays, and
the free-streaming scale is equal to \citep{2014NJPh...16f5002L}:
\begin{equation}
    k_{FS} = 0.82 \frac{\sqrt{\Omega_{\Lambda
          0}+\Omega_{m0}(1+z)^3}}{(1+z)^2}\frac{m_{\nu}}{1\mathrm{eV}}
    \ h \ \mathrm{Mpc}^{-1} \ ,
\end{equation}
where $h\equiv H_0/(100 \ \mathrm{km} \ \mathrm{s}^{-1}
\mathrm{Mpc}^{-1})$ is the dimensionless Hubble parameter,
$\Omega_{\Lambda 0}$ and $\Omega_{m0}$ are the cosmological constant
and the matter density parameters, respectively, evaluated at $z=0$,
and $m_{\nu}$ is the neutrino mass.  The physical effect of
free-streaming is to damp neutrino density fluctuations on scales $k
\gg k_{FS}$, where neutrinos cannot cluster due to their large
thermal velocity.  This affects the matter power spectrum since
neutrinos do not contribute, for $k>>k_{FS}$, to the gravitational potential wells
produced by dark matter and baryons. Hence the power spectrum is
reduced by a factor $\sim (1-f_{\nu})^2$, where
\begin{equation}
    f_{\nu} \equiv \frac{\Omega_{\nu}}{\Omega_m}
\end{equation}
is the neutrino mass fraction. For the same reason, the growth rate
of dark matter perturbations is suppressed and acquires a scale
dependence \citep{2008PhRvD..77f3005K}.

The neutrino mass has non-trivial effects also on the cosmic microwave
background (CMB) temperature anisotropies altering the redshift of
matter-radiation equality, if $\Omega_m h^2$ is kept fixed. This
translates into an overall modification of the amplitude and the
location of the acoustic peaks.  A change in the matter density would
instead affect the angular diameter distance to the last scattering
surface $D_A(z_{dec})$, and the slope of the CMB spectrum at low
multipoles, due to the Integrated Sachs-Wolfe effect
(\citealt{1967ApJ...147...73S,1985SvAL...11...271K}).  Many works
attempted to measure neutrino mass combining different cosmological
probes (e.g \citealt{2005PhRvL..95a1302W, 2006JCAP...10..014S,
  2009ApJS..180..306D, 2009PhRvD..79b3520I, 2010JCAP...01..003R,
  2010PhRvL.105c1301T, 2011ApJS..192...18K, 2011PhRvD..83d3529S,
  2012MNRAS.425..415S, 2009ApJS..180..225H, 2013ApJS..208...19H}).
One of the latest constraints come from recent Planck results
\citep{2015arXiv150201589P}, that put an upper limit on the sum of
neutrino masses, $\sum m_{\nu} < 0.23\ \mathrm{eV} $. Using instead
large scale structure probes, \cite{2014MNRAS.444.3501B} find that
$\sum m_{\nu} = 0.36 \pm 0.14\ \mathrm{eV}$, combining measurements
from the Baryon Oscillation Spectroscopic Survey (BOSS) CMASS DR11
with WMAP9. So they exclude massless neutrinos at $2.6\sigma$, and
including weak lensing and baryon acoustic oscillations (BAO)
measurements the significance is increased to $3.3\sigma$.

Among large scale structure probes, redshift-space distortions (RSD)
are one of the most promising ways to measure the neutrino mass.  RSD
are caused by galaxy peculiar velocities.  When galaxy distances are
computed from redshift measurements, assuming that the total velocity
relative to the observer comes only from the Hubble flow, one obtains
a distorted density field.  This distortion effect is clearly
imprinted in the two-point correlation function of galaxies. In
particular, the iso-correlation contours appear squashed along the
line of sight (LOS) on linear scales, while non-linear motions produce an
elongation effect known as Fingers of God.  The distortions on linear
scales can be quantified by the distortion parameter
\begin{equation}
    \beta(z) \equiv f(z)/b(z) \ , 
\end {equation}
which is the ratio of the growth rate of structures and their linear
bias factor.  The parameter $\beta(z)$ is strictly related to the
matter density parameter, since $f(z) = \Omega_m^{\gamma}(z)$, where
$\gamma$ is the linear growth factor \citep{2005PhRvD..72d3529L}.
Therefore, RSD provide the possibility to recover some important
information about the dynamics of galaxies and the amount of matter in
the Universe.

Massive neutrinos strongly affect the spatial clustering of cosmic
structures: as shown, for instance, in \cite{2011MNRAS.418..346M},
when assuming the same amplitude of primordial scalar perturbations,
the average number density of large scale structures (LSS) is
suppressed in the massive neutrino scenario, and the halo bias is
enhanced with respect to the massless case. Moreover, the value of
$f(z)$ decreases in the presence of massive neutrinos, due to their
free-streaming which suppresses structure formation. Therefore, the
value of $\beta$, which describes the cumulative effect of non-linear
motions, is reduced by an amount that increases with $\sum m_{\nu}$
and $z$. Moreover, free-streaming massive neutrinos induce also a
scale dependence in the parameter $\beta$. Finally, also the rms of
the galaxy peculiar velocity is reduced with respect to the massless
case, since both the growth rate $f(k,z)$ and the matter power
spectrum enter the bulk flow predicted by linear theory
(\citealt{2005NJPh....7...61E,2008PhRvD..77f3005K}).

At intermediate scales ($5\lesssim r[\mathrm{Mpc}/h] \lesssim 100$)
and low redshifts, these effects are degenerate with the amplitude of
the matter power spectrum, parameterised by $\sigma_8$.  Indeed, the
differences between the values of $\beta$ in a $\Lambda\mathrm{CDM}$
and a $\Lambda\mathrm{CDM}+\nu$ models are significantly reduced if
the two cosmologies are normalised to the same value of $\sigma_8$.
Nonetheless, the relative difference between the theoretical values of
$\beta$ in these two models, at $z=1$, is $\delta\beta/\beta \simeq
3\%$, for $\sum{m_{\nu}}=0.6\ \mathrm{eV} $, which corresponds to the
precision reachable by future redshift surveys in measuring the
redshift space distortion parameter at $z < 1$
\citep{2011MNRAS.418..346M}.  RSD can thus contribute to constrain the
total neutrino mass, helping to disentangle the degeneracies with
other cosmological parameters.

The aim of this work is to exploit RSD to constrain cosmological
parameters through a Markov Chain Monte Carlo (MCMC) procedure and
make forecasts on the statistical accuracy achievable with future
cosmological probes.  Some attempts have been recently made to produce
forecasts based on RSD using numerical simulations.  For example,
\cite{{2008Natur.451..541G}} used mock surveys extracted from the
Millennium simulation to estimate the errors affecting measurements of
the growth rate.  They found a scaling relation for the relative error
on the $\beta$ parameter as a function of the survey volume and mean
density.  This formula has been later refined by
\cite{2012MNRAS.427.2420B}.  The authors analysed the same catalogues
of dark matter haloes used in the present work, extracted from a
snapshot of the {\small BASICC} simulation \citep{2008MNRAS.383..755A}
at $z = 1$, finding that the parameter $\beta$ can be underestimated
by up to $10\%$, depending on the minimum mass of the considered
haloes.  They also proposed a new fitting formula, that aims at
separating the dependence of the statistical error on bias, density
and volume:
\begin{equation}
    \frac{\delta\beta}{\beta} \approx C b^{0.7} V^{-0.5} \exp \left(
    \frac{n_0}{b^2n} \right) \ ,
    \label{bianchi_form}
\end{equation}
where $n_0 = 1.7 \cdot 10^{-4} \, h^{3} \mathrm{Mpc}^{-3}$ and $C =
4.9 \cdot 10^{2} \, h^{-1.5} \mathrm{Mpc}^{1.5}$.

Here we follow a similar approach to study how the error on
cosmological parameters depends on the survey parameters, focusing in particular on the neutrino mass
fraction.  The main differences with respect to the work of \cite{2012MNRAS.427.2420B} are the following:
\begin{itemize}
\item we use a theoretical real-space correlation function obtained from the dark matter power spectrum instead of the deprojected one;
\item we use the multipoles of the correlation function rather than the full two-dimensional correlation function;
\item we use an MCMC likelihood analysis to estimate parameters.
\end{itemize}
The combination of monopole and quadrupole is fundamental to break the
degeneracy between the halo bias and $f_{\nu}$, and thus to constrain
the neutrino mass fraction, as we will discuss later in detail.

This paper is organised as follows. In \S \ref{sec BASICC} we describe
the BASICC simulation and the method adopted to select the subsamples.
In \S \ref{sec method} we describe the modellisation of the
correlation function, the construction of the covariance matrix, and
the approach used for the estimation of the best-fit parameters.  In
\S \ref{sec res} we present our results, showing the dependence of the
errors on the simulation parameters, providing a fitting formula
similar to Eq.~(\ref{bianchi_form}).  Finally, in \S \ref{sec concl} we
draw our conclusions.


\section{Halo catalogues from the BASICC simulation}
\label{sec BASICC} 
One of the building blocks of our work is the {\small BASICC}
simulation, the Barionic Acoustic oscillation Simulation produced at the
Institute for Computational Cosmology \citep{2008MNRAS.383..755A}.
One of the advantages of using numerical simulations is that we know a
priori the value of the parameters we want to measure.  Moreover,
simulations solve the problem of having only one Universe available
for observations. Indeed it is possible to construct many mock
catalogues, assuming the same cosmological parameters, and repeat the
measurements for each of them.  In particular, comparing the
theoretical values of the parameters we want to measure with the mean
of their measured estimates, we can assess the systematic errors due
to the method, while the scatter between measurements gives us an
estimate of the expected statistical errors.

The {\small BASICC} simulation has been explicitly designed to study
BAO features in the clustering pattern, so its volume is large enough
to follow the growth of fluctuations on a wide range of scales. At the
same time, its mass resolution is high enough to allow splitting the
whole box in sub-cubes with the typical volumes of ongoing surveys,
still preserving a good statistics on the scales which are central in
the present analysis.  The BASICC simulation is made up by $1448^3$
dark matter particles of mass $M_{part} = 5.49 \times 10^{10} \,
h^{-1} \, M_\odot$, in a periodic box of side $1340 \, h^{-1}
\text{Mpc}$.  The cosmological model adopted is a $\Lambda CDM$ model
with $\Omega_m = 0.25$, $\Omega_\Lambda = 0.75$, $\sigma_8 = 0.9$ and
$h = H_0 / (100 \, \text{km} \, \text{s}^{-1} \, \text{Mpc}^{-1}) =
0.73$.  The dark matter haloes are identified using a
Friends-of-Friends (FOF) algorithm \citep{1985ApJ...292..371D} with a
linking length of $0.2$ times the mean particle separation.  We
consider only haloes with a minimum number of particles per halo of
$N_{part} = 20$, so that the minimum halo mass is $M_{halo} = 20 \cdot
M_{part} \simeq 1.1 \times 10^{12} \ h^{-1} \ M_\odot$.

\begin{table}
  \caption{Properties of the halo catalogues used in the
    analysis. $N_{part}$ is the minimum number of particles per halo;
    $M_{cut}$ is the corresponding threshold mass; $\mathcal{N}_{tot}$
    is the number of haloes with $M_{halo} \ge M_{cut}$; $n$ is the
    number density, computed as $\mathcal{N}_{tot}/V$, where $V =
    (1340 \ h^{-1} \mathrm{Mpc})^3$ is the simulation volume; $b$ is the
    bias value.}
  \begin{center}
    \begin{tabular}{ccccc}
      \toprule
      $N_{part}$ & $M_{cut}\times 10^{-12}$ & $\mathcal{N}_{tot}$ & $n\times 10^5$ & $b$\\
       & $[h^{-1} \ M_\odot]$ & & $[h^3 \ \text{Mpc}^{-3}]$ &  \\
      \midrule
      $20$ & $1.10$ & 7483318 & $311$ & $1.44$\\
      $63$ & $3.46$ & 2164960 & $90.0$ & $1.80$\\
      $136$ & $7.47$ & 866034 & $36.0$ & $2.15$\\
      $236$ & $13.0$ & 423511 & $17.6$ & $2.49$\\
      $364$ & $20.0$ & 230401 & $9.58$ & $2.89$\\
      \bottomrule
    \end{tabular}
  \end{center}
 \label{tab halo masses}
\end{table}
\begin{table}
  \caption{Sub-samples used in our analysis to explore the dependence
    of the errors on mean density, bias and volume. Each sample is
    characterised by given values of the mean density, $n$, and the mass
    threshold, $M_{cut}$, (or the bias, $b$). The full, non-diluted, samples
    coincide with the bottom entry of each column. The entries in the
    table identified by circles represent the samples used to test the
    dependence of the errors on the survey volume. For these samples the
    simulation box has been split in $N^3$ sub-boxes with
    $N=\{4,5,6\}$, whereas for the other sub-samples (asterisks) we
    only consider $N=3$.}
  \begin{center}
    \begin{tabular}{cccccccc}
      \toprule
      & & \multicolumn{5}{c}{$M_{cut}\times 10^{-12} \ [h^{-1} \ M_{\odot}]$} \\
      & & $1.10$  & $3.46$ & $7.47$ & $13.0$ & $20.0$ \\
      \midrule
      & 6.87 & $\ast$ & $\ast$ & $\ast$ & $\ast$ & $\ast$ \\
      & 9.58 & $\ast$ & $\ast$ & $\ast$ & $\ast$ & $\ast$ \\
      & 12.1 & $\ast$ & $\ast$ & $\ast$ & $\ast$ & \\
      & 17.6 & $\ast$ & $\ast$ & $\ast$ & $\ast$ & \\
      & 24.8 & $\circ$ & $\circ$ & $\circ$ & & \\
      $n \times 10^5$ & 36.0 & $\ast$ & $\ast$ & $\ast$ & & \\
      $[h^{3} \, \text{Mpc}^{-3}]$ & 58.7 & $\ast$ & $\ast$ & & & \\
      & 90.0 & $\circ$ & $\ast$ & & & \\
      & 131 & $\ast$ & & & & \\
      & 204 & $\ast$ & & & & \\
      & 311 & $\circ$ & & & & \\
      \bottomrule
    \end{tabular}
  \end{center} 
  \label{tab sub-samples}
\end{table}

In the present work, we consider the snapshot at $z=1$, that is the central value in the
range of redshifts that will be explored in future redshift surveys,
and select halo catalogues with different mass thresholds
(i.e. different minimum number of particles per halo), which means
different bias values.  The properties of these catalogues are
summarised in Table \ref{tab halo masses}.  This selection allows us
to study the dependence of the error on the sample bias.  Moreover, in
order to investigate also the dependence of the errors on the halo density, the
samples have been diluted to create a series of catalogues with
decreasing density, down to a value of $\sim 7 \times 10^{-5} h^3
\ \text{Mpc}^{-3}$, at which the shot noise starts to dominate (see
\citealt{2012MNRAS.427.2420B}).  For each of these samples with
varying bias and density, we split the simulation box in $3^3$
sub-boxes, obtaining $27$ sub-boxes.  For some samples we also split
the box in $N^3$ parts with $N=\{4,5,6\}$, in order to explore the
error dependence on the volume, as shown in Table \ref{tab sub-samples}.

\begin{figure*}
  \begin{center}
    \subfigure{\includegraphics[width=8cm]{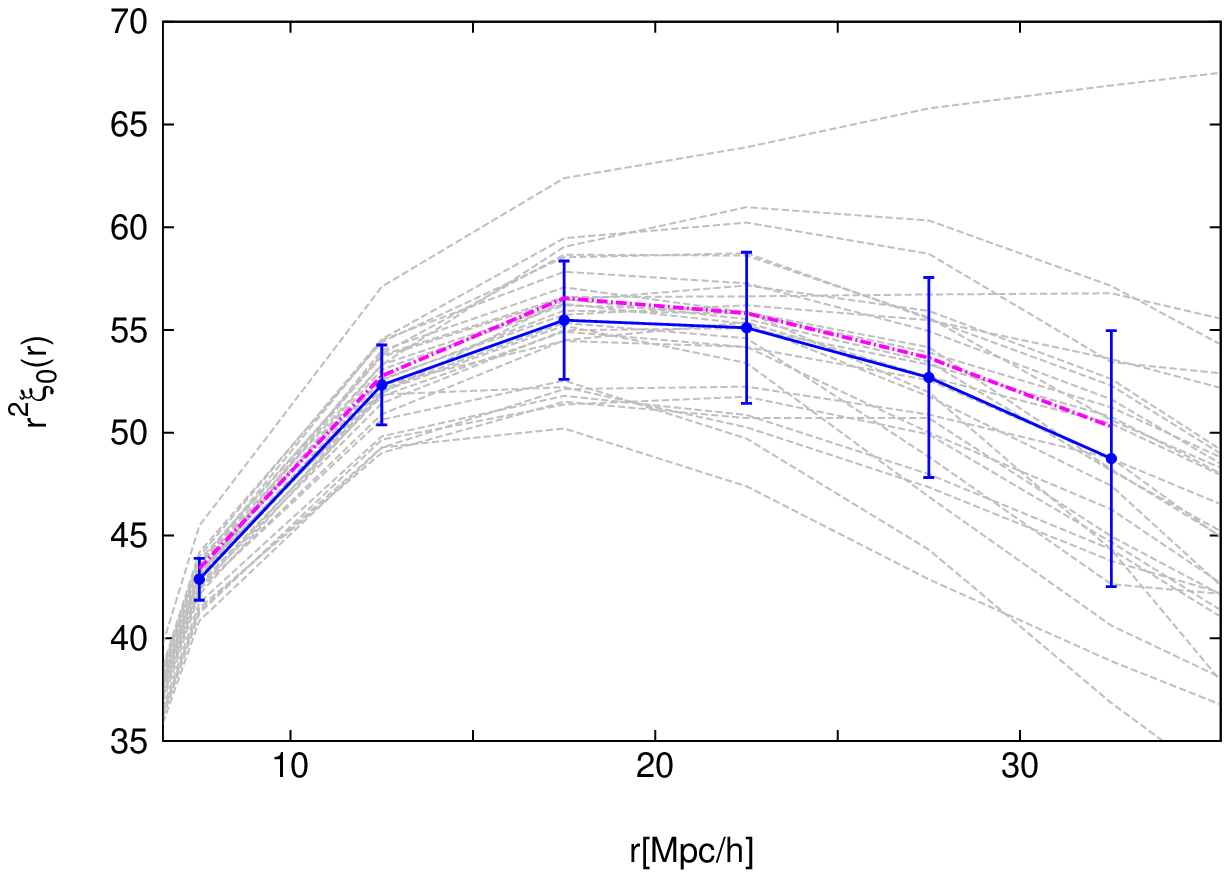}}
    \subfigure{\includegraphics[width=8cm]{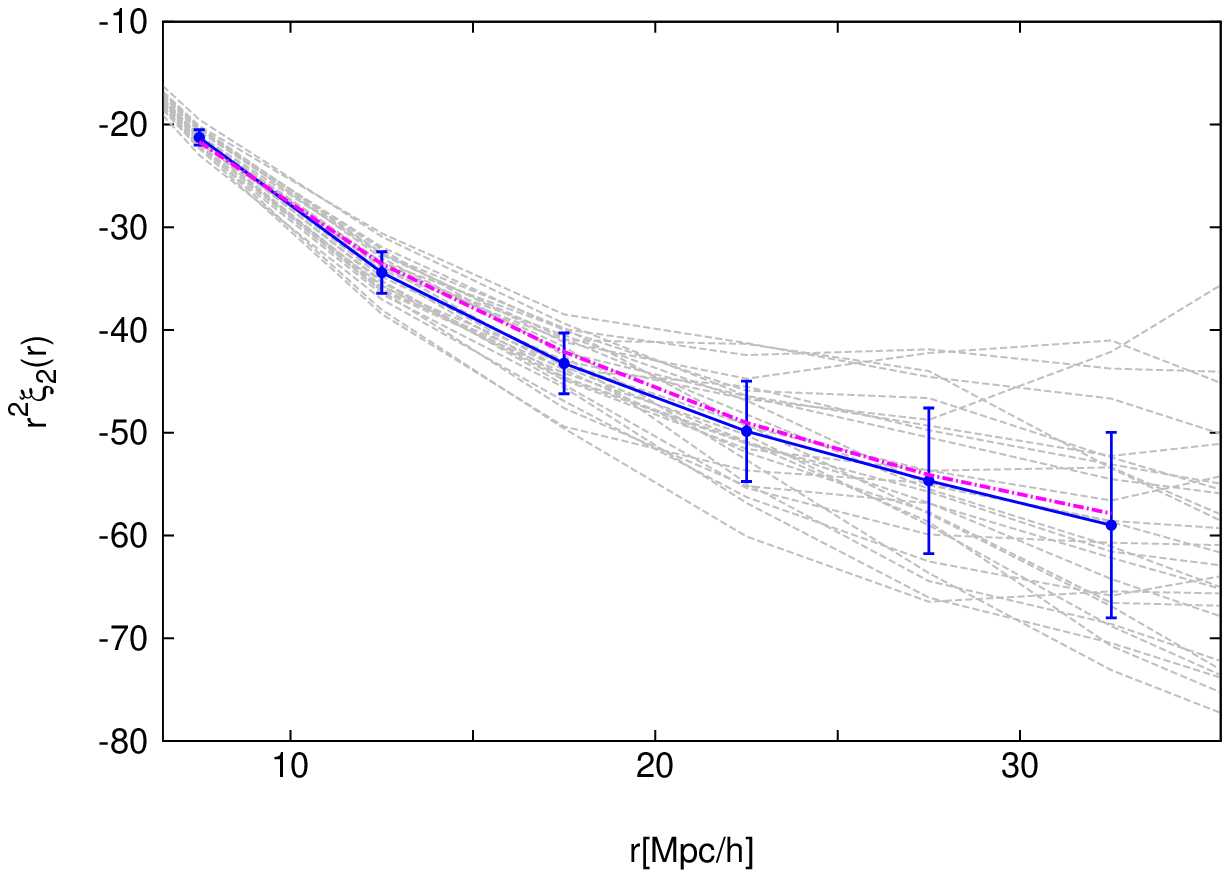}}
    \caption{Monopole $\xi_0$ (left panel) and quadrupole $\xi_2$
      (right panel) of the correlation function, multiplied by
      $r^2$. The grey dashed lines represent the multipoles measured
      from the $27$ mocks extracted from the catalogue with a mass
      threshold of $M_{cut} = 1.1\times10^{12} \ h^{-1}
      \ M_{\odot}$. The blue solid lines represent the multipoles
      averaged over the $27$ mocks, the error bars being the square
      root of the diagonal elements of the covariance matrix. The
      best-fit mean model is represented by the magenta dot-dashed
      lines.}
    \label{multi02}
  \end{center}
\end{figure*}
\section{Methodology}
\label{sec method}
In this section we describe the method adopted to measure the correlation
function from the mock catalogues, the modellisation of the
correlation function and its multiploes, and the computation of the
covariance matrix needed for the likelihood analyses.

\subsection{Correlation function measurement}
\label{sec corr-func-sim}
The two-dimensional two-point correlation function has been evaluated
using the Landy \& Szalay (1993) estimator:
\begin{equation} 
  \label{eq LS}
    \xi(r_p,\pi) = \frac{HH(r_p,\pi)-2HR(r_p,\pi)+RR(r_p,\pi)}{RR(r_p,\pi)} \; ,
\end{equation}
where $r_p$ and $\pi$ are, respectively, the separation perpendicular
and parallel to the LOS, that is defined as the direction from the
observer to the centre of each pair.  The quantities $HH$, $HR$ and
$RR$ represent the normalised halo-halo, halo-random and random-random
pair counts at a given distance range, respectively.  The random
catalogues have 50 times the number of objects of the mock
catalogues\footnote{To measure the two-point correlation functions we
  make use of the CosmoBolognaLib \citep{2015arXiv151100012M}, a large
  set of Open Source C++ libraries freely available at this link:
  http://apps.difa.unibo.it/files/people/federico.marulli3/}. The bin
size used to compute the two-dimensional correlation function is $1 \,
\mathrm{Mpc}\, h^{-1} \times\, 1 \, \mathrm{Mpc}\, h^{-1}$, and the
maximum separation considered in the pair counts is $s =
\sqrt{{r_p}^2+\pi^2} = 50 \, \mathrm{Mpc} \, h^{-1} $.

The multipoles are then computed in bins of $5 \,
\mathrm{Mpc}\,h^{-1}$, integrating the two-dimensional correlation
function as follows:
\begin{equation}
    \begin{split}
    \label{multipoles}
    \xi_l(r) & = \frac{2l+1}{2} \int^1_{-1} \xi(r_p,\pi) P_l(\mu) d\mu \\
          & = \frac{2l+1}{2} \int^{\pi}_0 \sqrt{1-\mu^2} \xi(r_p,\pi) P_l(\mu) d\theta \; ,
    \end{split}
\end{equation}
where $P_l(\mu)$ are the Legendre polynomials and $\mu$ is the cosine
of the angle between the separation vector and the LOS: $\mu =
\cos\theta = \pi/r_p$.  In this work we will consider only the
monopole and the quadrupole, where the most relevant information is
contained, and ignore the contribution of the noisier subsequent
orders.

\begin{figure*}
  \begin{center}
    \subfigure{\includegraphics[width=5.8cm, trim = 35 0 30 30]{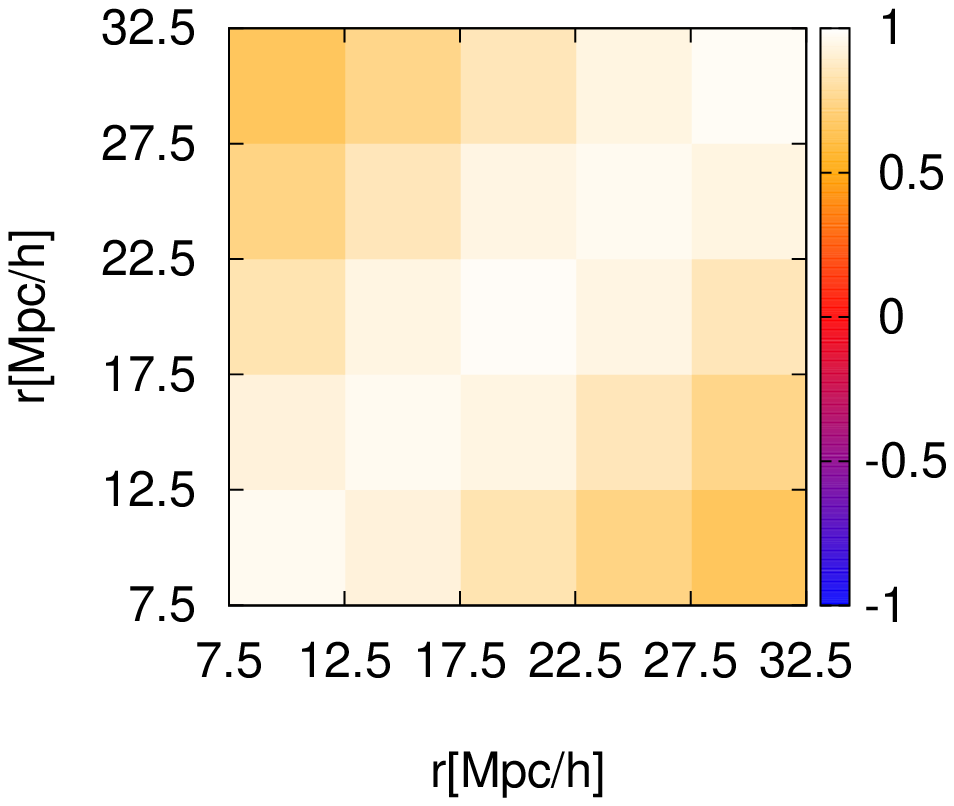}}
    \subfigure{\includegraphics[width=5.8cm, trim = 32.5 0 30 32.5]{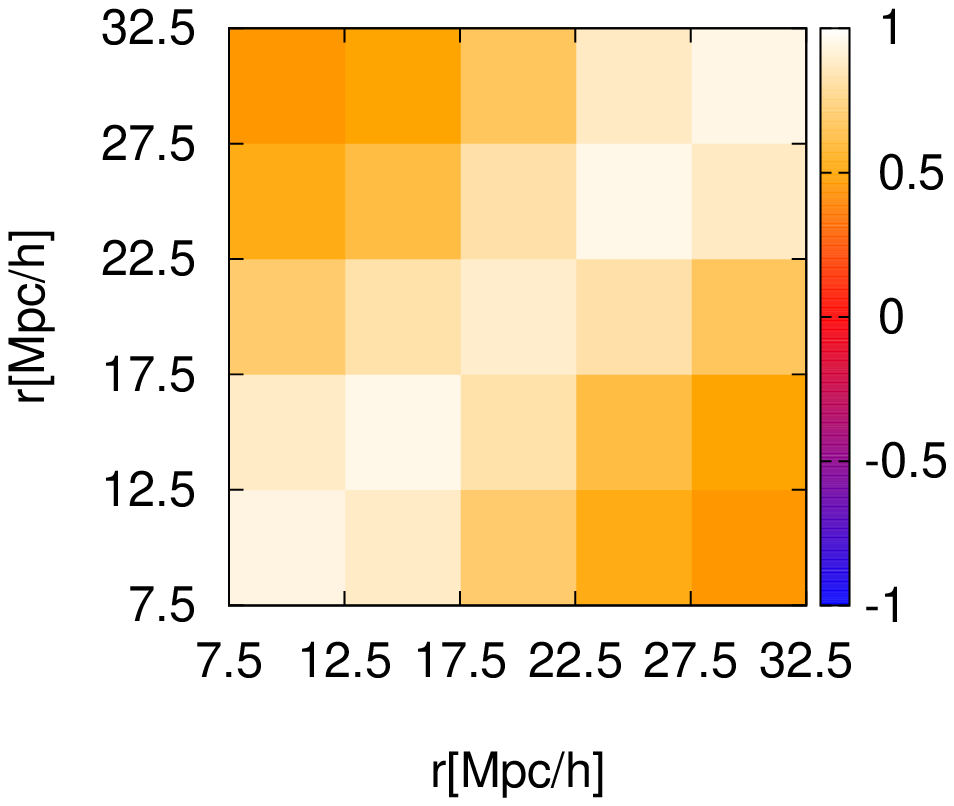}}
    \subfigure{\includegraphics[width=5.8cm, trim = 30 0 30 35]{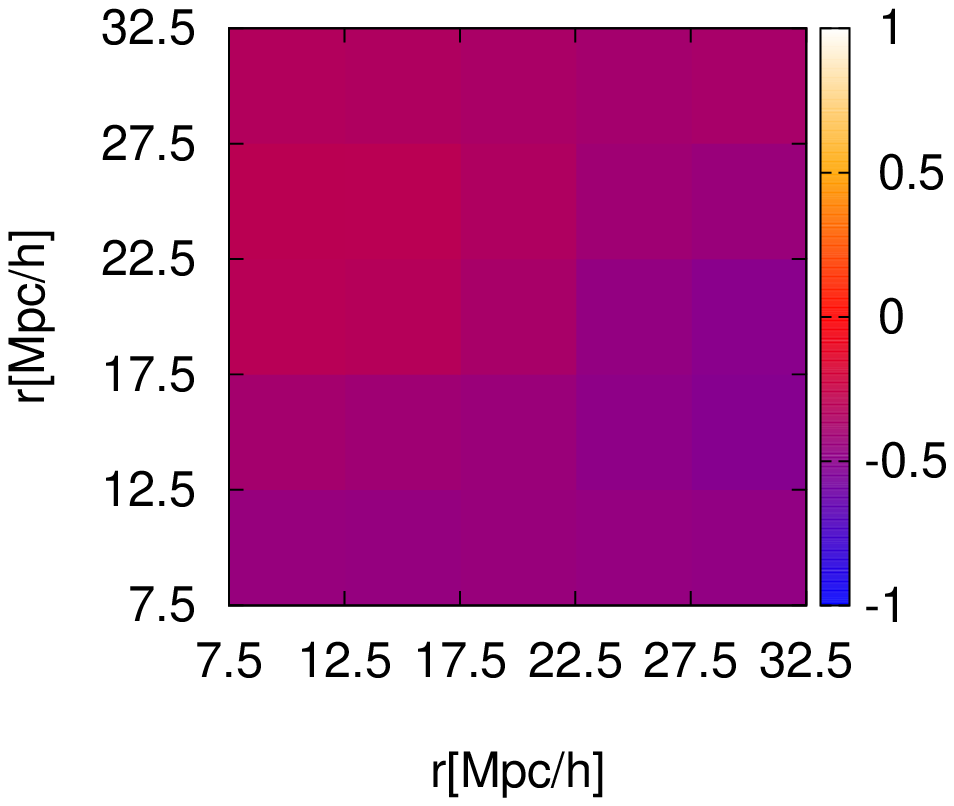}}
  \end{center}
  \caption{Reduced covariance matrix constructed for the monopole
    (left panel) and quadrupole (central panel) and the cross
    covariance between the two (right panel) in bins of $5
    \ \mathrm{Mpc} \ h^{-1}$, computed from the $27$ mock catalogues
    of the most dense sample with mass threshold $M_{cut} = 1.1
    \times10^{12} \ h^{-1} \ M_{\odot} $}.
    \label{covmat_02}
\end{figure*}

\subsection[]{Correlation function model} \label{sec corr-func-mod}
We compute the non-linear power spectrum, $P_{nl}(k)$, at $z=1$ using
CAMB \citep{2000ApJ...538..473L}, for different values of $f_{\nu}
\equiv \Omega_{\nu}/\Omega_m$.
Then the theoretical real-space
correlation function $\xi(r)$ is obtained by Fourier transforming the
non-linear power spectrum.  As pointed out by
\citet{1987MNRAS.227....1K} and later by \citet{1992ApJ...385L...5H},
in the linear regime (i.e. at sufficiently large scales) and in the
plane-parallel approximation, the two-dimensional correlation function
in redshift-space can be written as:
\begin{equation}
    \xi^*(r_p,\pi) = \xi_0(s) P_0(\mu) + \xi_2(s) P_2(\mu) + \xi_4(s) P_4(\mu) \; .
    \label{eq lin model}
\end{equation}
The multipole moments $\xi_l(s)$ of the correlation function are defined as:  
\begin{eqnarray}
    \xi_0(r) & \equiv & \bigg(1 + \frac{2}{3}\beta + \frac{1}{5}\beta^2 \bigg) \xi(r) \; , \\
    \xi_2(r) & \equiv & \bigg(\frac{4}{3}\beta + \frac{4}{7}\beta^2 \bigg) [\xi(r) - \bar{\xi}(r)] \; , \\
    \xi_4(r) & \equiv & \frac{8}{35}\beta^2 \bigg[\xi(r) + \frac{5}{2}\bar{\xi}(r) - \frac{7}{2}\bar{\bar{\xi}}(r) \bigg] \; ,
    \label{eq mult}
\end{eqnarray}
where $\beta$ is the redshift-space distortion parameter that
describes the squashing effect on the iso-correlation contours in
redshift space along the direction parallel to the LOS; $\xi(r)$ is
the real-space undistorted correlation function, while $\bar{\xi}$ and
$\bar{\bar{\xi}}$ are defined as:
\begin{eqnarray}
    \label{eq xi bar}
    \bar{\xi} & \equiv & \frac{3}{r^3} \int_0^r \xi(r') r'^2 dr' \; , \\
    \label{eq xi barbar}
    \bar{\bar{\xi}} & \equiv & \frac{5}{r^5} \int_0^r \xi(r') r'^4 dr' \; .
\end{eqnarray}

This model describes the RSD only at large scales, where non-linear
effects can be neglected.  In order to take into account the
non-linear dynamics, we convolve the linearly distorted redshift-space
correlation function with the distribution function of random pairwise
velocities along the LOS, $f(v)$:
\begin{equation}
    \label{eq lin-exp model}
    \xi(r_p,\pi) = \int_{-\infty}^{+\infty}
    \xi^*\biggl[r_p,\pi-\dfrac{v(1+z)}{H(z)}\biggr] f(v) dv \; .
\end{equation}
The distribution function $f(v)$ is a function that represents the
random motions and can be expressed by a Gaussian form:
\begin{equation}
    \label{vel dist Gauss}
    f(v) = \dfrac{1}{\sigma_{12} \sqrt{\pi}} \exp \bigg(-
    \dfrac{v^2}{\sigma^2_{12}} \bigg)
\end{equation}
\citep{1983ApJ...267..465D, 1994MNRAS.267..927F,Peacock}, where
$\sigma_{12}$ does not depend on pair separations and can be
interpreted as the pairwise velocity dispersion.

The non-linear model given by Eq.~(\ref{eq lin-exp model}) is then
integrated to obtain the multipoles according to
Eq.~(\ref{multipoles}).  So the multipoles of both the measured and
the theoretical correlation functions are computed in the same way
using the measured correlation function, Eq.~(\ref{eq LS}), and the
model correlation function, Eq.~(\ref{eq lin-exp model}),
respectively, thus minimising any numerical bias.  As an example,
Fig.~\ref{multi02} shows the comparison between the multipoles
computed from the $27$ mock catalogues extracted from the most dense
sample with a mass threshold of $M_{cut} = 1.1 \times10^{12} \ h^{-1}
\ M_{\odot} $, and their best-fit model.  We can appreciate the
agreement between the model (magenta dot-dashed lines), obtained by
fixing all parameters to their best-fit values, and the mean
multipoles (blue dots) computed over the $27$ mock catalogues (grey
dashed lines). The mean difference between the two is $\sim 5\%$ for
the monopole and $\sim 13\%$ for the quadupole.

\subsection{Covariance Matrix and Likelihood}
We use the $27$ mock catalogues extracted from the {\small BASICC}
simulation to estimate the covariance matrix.  We compute the
multipoles of the correlation function for each mock catalogue and
construct the covariance matrix as follows:
\begin{equation}
    C_{ij} = \frac{1}{N-1}\sum^N_{k=1}(\bar{X}_i-X^k_i)(\bar{X}_j-X^k_j) \; ,
    \label{covmat}
\end{equation}
where the sum is over the number of mocks $N=27$, and $\mathbf{X}$ is the
data vector containing the multipole vectors:
\begin{equation}
  \begin{split}
    \mathbf{X} = & \{\xi^{(1)}_0,\xi^{(2)}_0,...,\xi^{(M)}_0,
    \xi^{(1)}_2,\xi^{(2)}_2,...,\xi^{(M)}_2\} \; ,
  \end{split}
\end{equation}
with $M$ being the number of bins, {\emph i.e.} the dimension of each
multipole vector.  In particular, $\bar{X}_i$ is the mean value over
the $27$ catalogues of the $i^{th}$ element of the data vector, while
$X^k_i$ is the value of the $i^{th}$ component of the vector
corresponding to the $k^{th}$ mock catalogue.  Fig.~\ref{covmat_02}
shows the reduced covariance matrix defined as $\tilde{C}_{i,j} =
C_{i,j}/\sqrt{C_{i,i}C_{j,j}}$. We can see that there are significant
off-diagonal terms, and a non-negligible covariance between monopole
and quadrupole.  However, in this work we are going to consider only
the diagonal part of the matrix, since this simplification does not
affect our final results and reduces numerical noises (see Appendix
\ref{app} for details).

The likelihood is assumed to be proportional to $\exp(-\chi^2/2)$
\citep{NR3}, where $\chi^2$ is defined as:
\begin{equation}
    \label{eq chi2}
    \chi^2 \equiv \sum^{N_{bins}}_{i,j=1}(X_{th,i}-X_{obs,i})C^{-1}_{ij}(X_{th,j}-X_{obs,j}) \; ;
\end{equation}
$N_{bins}$ is the length of the vector X, which is twice the length of
each multipole vector. $X_{th}$ is the multipole vector computed from
the theoretical correlation function and $X_{obs}$ is the data vector
computed from the simulation for each catalogue of Table \ref{tab
  sub-samples}.

\begin{figure}
    \begin{center}
        \includegraphics[width=0.45\textwidth, trim = 20 0 20 0 ]{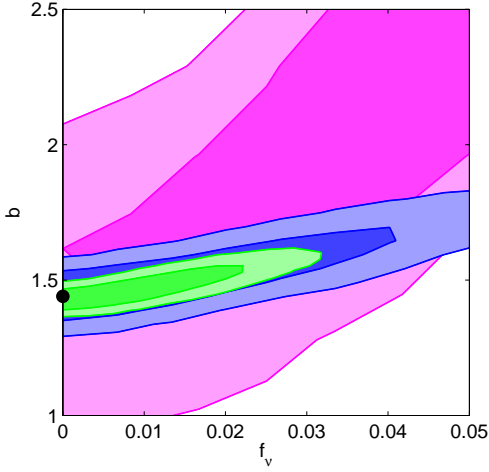}
    \end{center}
    \caption{Contour plots for $f_\nu$ and $b$ derived using the
      monopole only (blue contours), the quadrupole only (magenta
      contours) and from their combination (green contours). The
      results have been obtained from one of the mock catalogues with
      mass $M_{cut} = 1.1\times10^{12} M_\odot \ h^{-1} $ and density
      $n = 3.1\times10^{-3} h^3 \ \mathrm{Mpc}^{-3}$. The input values
      of the simulation, represented by the black dot, are $f_{\nu} =
      0$ and $b=1.44$. Dark and light ellipses represent $1\sigma$ and
      $2\sigma$ contours, respectively.}
    \label{bfnu_best_comp}
\end{figure}
\begin{figure}
  \begin{center}
    \includegraphics[width=0.45\textwidth, trim = 20 0 20 0 ]{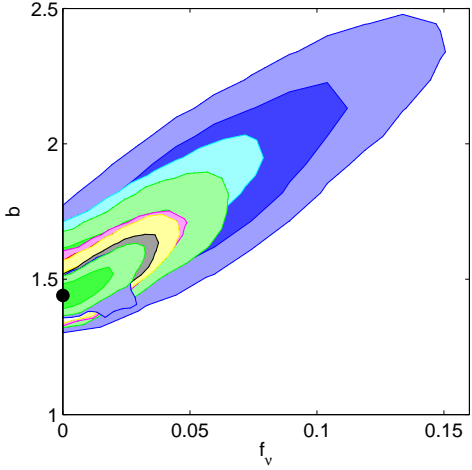}
  \end{center}
  \caption{Contour plots for $f_{\nu}$ and $b$ for a single mock
    catalogue from the samples with $M_{cut} = 1.1\times10^{12}
    \ M_{\odot} \ h^{-1}$ and density values as in the first column of
    Table \ref{tab sub-samples}. Larger contours correspond to lower
    density samples. The input values of the simulation,
    highlighted by the black dot, are recovered within $1\sigma$.}
  \label{contour_bfnu}
\end{figure}
\begin{figure}
    \begin{center}
        \includegraphics[width=0.45\textwidth, trim = 20 0 20 0 ]{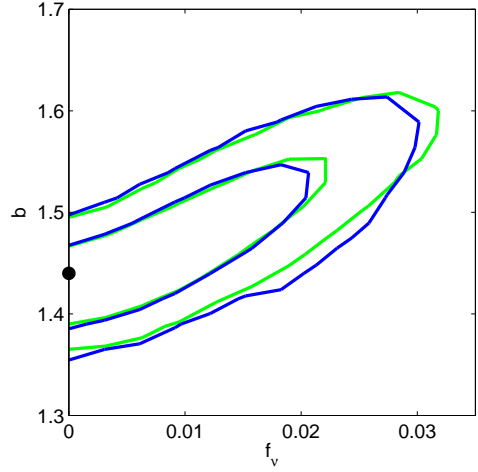}
    \end{center}
    \caption{Contour plots for $f_\nu$ and $b$ derived considering
      $f(z) = \Omega^{\gamma}_{m}(z)$ (green contours) and $f(z)
      \simeq [\Omega_{m}(z)(1-f_{\nu})]^{\gamma} $ (blue contours),
      using both monopole and quadrupole. The results have been
      obtained from one of the mock catalogues with mass $M_{cut} =
      1.1\times10^{12} M_\odot \ h^{-1} $ and density $n =
      3.1\times10^{-3} h^3 \ \mathrm{Mpc}^{-3}$.}
    \label{bfnu_growth_comp}
\end{figure}

\subsection{MCMC analysis} \label{MCMC}
We analyse the mock data with a MCMC procedure.  We explore a
three-dimensional parameter space considering the neutrino mass
fraction $f_{\nu} \equiv \Omega_{\nu}/\Omega_{m}$, the halo bias
parameter, $b$, and the pairwise velocity dispersion, $\sigma_{12}$.
The other cosmological parameters are kept fixed to the input values
of the simulations. To investigate the impact of this assumption, we
repeated our analysis assuming Plank-like priors for $\Omega_{\rm
  dm}h^2$, $\Omega_{b}h^2$ and $A_s$. Specifically, we allowed each of
these parameters to vary in the ranges $\Delta\Omega_{m}h^2 = 0.001$,
$\Delta\Omega_{b}h^2 = 0.00014$, and $\Delta\ln(10^{10}A_s) = 0.023$
\citep{2015arXiv150201589P}, around the input values of the
simulation. As we verified, the effect on our final results is
negligible, considering the estimated errors.

The neutrino mass fraction enters the model through the shape of the
real-space undistorted correlation function.  The bias instead enters
the model twice: first, when converting the real-space correlation
function of matter into the halo correlation function assuming a
linear biasing model, $\xi_{halo}(r) = b^2\xi_{m}(r) $, and second in
the multipole expansion through the parameter $\beta$, which in our
analysis is expressed as $\Omega^{\gamma}_{m}(z)/b(z)$, with $\gamma =
0.55$ according to \cite{2005PhRvD..72d3529L}.  $\Omega_m(z)$ is the
input value of the simulation computed at redshift $z=1$ via the
equation:
\begin{equation}
    \Omega_m(z) = \frac{{(1 + z)}^3 \Omega_{m0}}{{(1 + z)}^3
      \Omega_{m0} + (1 - \Omega_{m0})} \; .
\end{equation}

We assume the expression $f(z) = \Omega^{\gamma}_{m}(z)$, neglecting
the dependence on $f_{\nu}$ that the growth rate acquires at the
scales of interest in this work.  Nevertheless, we verified that this does not affect
significantly the results, as we will show in the next
section.

Once the theoretical correlation function is computed assuming a given
set of cosmological parameters, it should be rescaled to the fiducial
cosmology used to measure the correlation function, that in our case
is the input cosmology of the simulation.  This is done by adopting
the relation (e.g. see \citealt{2003ApJ...598..720S}):
\begin{equation}
    \xi^{fid}_{th}(r_p,\pi) = \xi_{th} \bigg( \frac{D_A(z)}{D^{fid}_A(z)}r_p,\frac{H^{fid}(z)}{H(z)}\pi \bigg) \; ,
\end{equation}
where $D_A(z)$ is the angular diameter distance and $H(z)$ is the
Hubble parameter, at redshift $z$.  However, in our case this
procedure is not necessary since the only varying cosmological
parameter is $f_{\nu}$, whereas the total amount of matter $\Omega_m$
is held fixed to the input value of the simulation, so that $H(z)$ and
$D_A(z)$ do not change and there are no geometric distortions to be
accounted for.

\section{Results} \label{sec res}
In this section we present our results.  First, we compare the
cosmological values recovered with the MCMC procedure with the input
values of the simulation.  Then we show how the errors on $f_{\nu}$
and bias depend on the halo density and the volume covered by the
simulation, and on the bias of the considered sample.

\begin{figure*}
  \begin{center}
      \includegraphics[width=16cm]{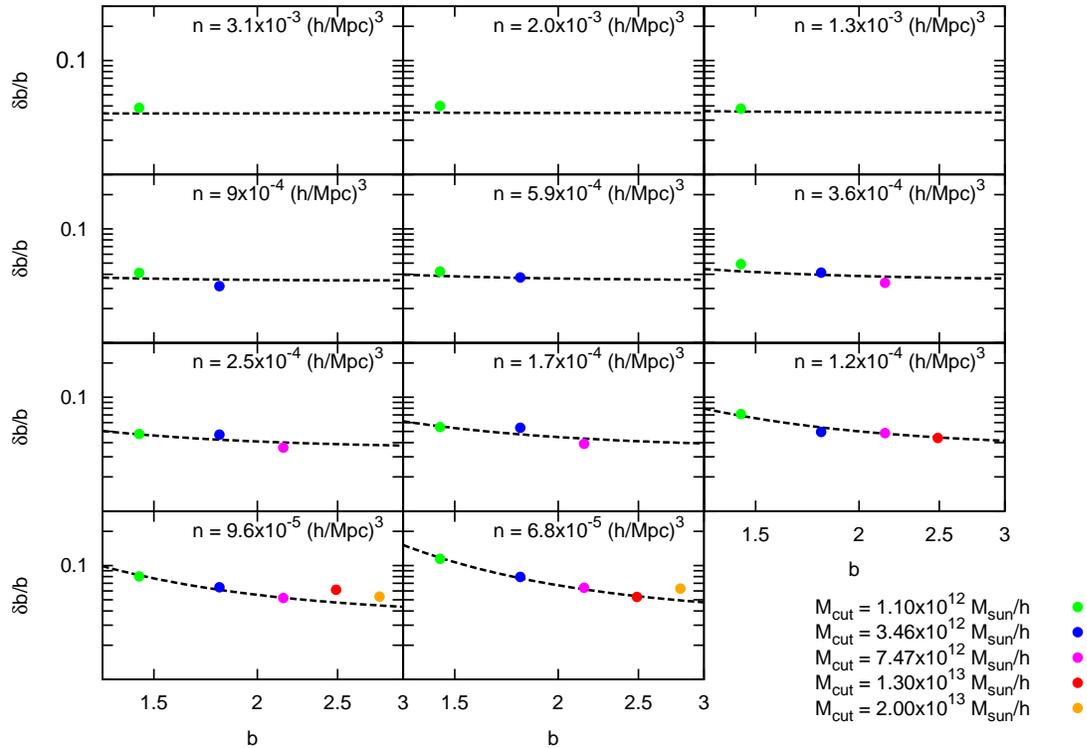}
  \end{center}
  \caption{Relative errors on the halo bias, $\delta b/b$, as a
    function of bias, $b$, for different mass (highlighted by
    different colours) and density samples, as labelled in the
    panels. The dots represent the mean error over the $27$ mock
    catalogues, the dashed lines show the scaling formula obtained by
    fitting our results, Eq.~(\ref{func}).}
  \label{bE_vs_bias}
\end{figure*}
\begin{figure*}
  \begin{center}
     \includegraphics[width=16cm]{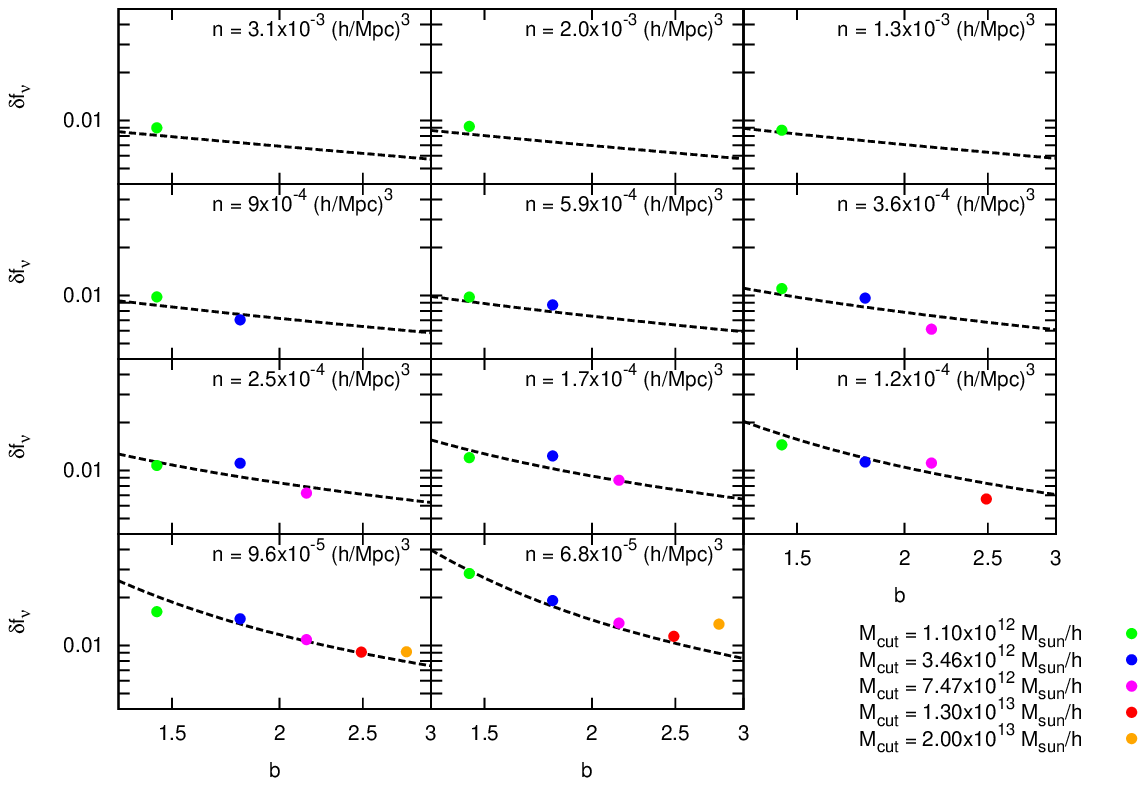}
  \end{center}
  \caption{As Fig.~\ref{bE_vs_bias} but for the errors on neutrino
    mass fraction, $\delta f_{\nu}$, as a function of bias, $b$, for
    different mass and density samples.}
  \label{fE_vs_bias} 
\end{figure*}

\subsection{Estimating the neutrino mass fraction} \label{sec res cont}
The joint constraints on the neutrino mass fraction, $f_{\nu}$, and
bias, $b$, marginalised over the pairwise velocity, $\sigma_{12}$, are
shown in Fig.~\ref{bfnu_best_comp}.  They have been obtained from one
mock catalogue of the most dense sample with a mass threshold of
$M_{cut} = 1.1\times10^{12} M_\odot \ h^{-1}$, using monopole and
quadrupole separately (blue and magenta contours, respectively), and
monopole and quadrupole together (green contours).  Let us notice that
the use of both monopole and quadrupole can significantly help to
tighten the constraints on both parameters.  Indeed, when modelling
only the monopole, there is a degeneracy between the halo bias and
$f_{\nu}$, since they affect the normalisation of $\xi_{halo}$ in
opposite directions.  On the other hand, the quadrupole moment, which
includes the effects of RSD, can help in breaking this degeneracy,
especially for large values of $f_\nu$.  Therefore, the combination of
the first two multipoles of the redshift-space two-point correlation
function is crucial to estimate the neutrino mass fraction.  We can
also notice that the input values of the two parameters, $f_{\nu} = 0$
and $b = 1.44$, are recovered within $1\sigma$ contours.

In Fig.~\ref{contour_bfnu} we show the $1\sigma$ and $2\sigma$
contours obtained using both monopole and quadrupole for the same mock
catalogue of the previous figure, but considering the different
density values reported in Table \ref{tab sub-samples}. Larger
contours correspond to catalogues with lower densities.

\begin{figure*}
  \begin{center}
       \includegraphics[width=16cm]{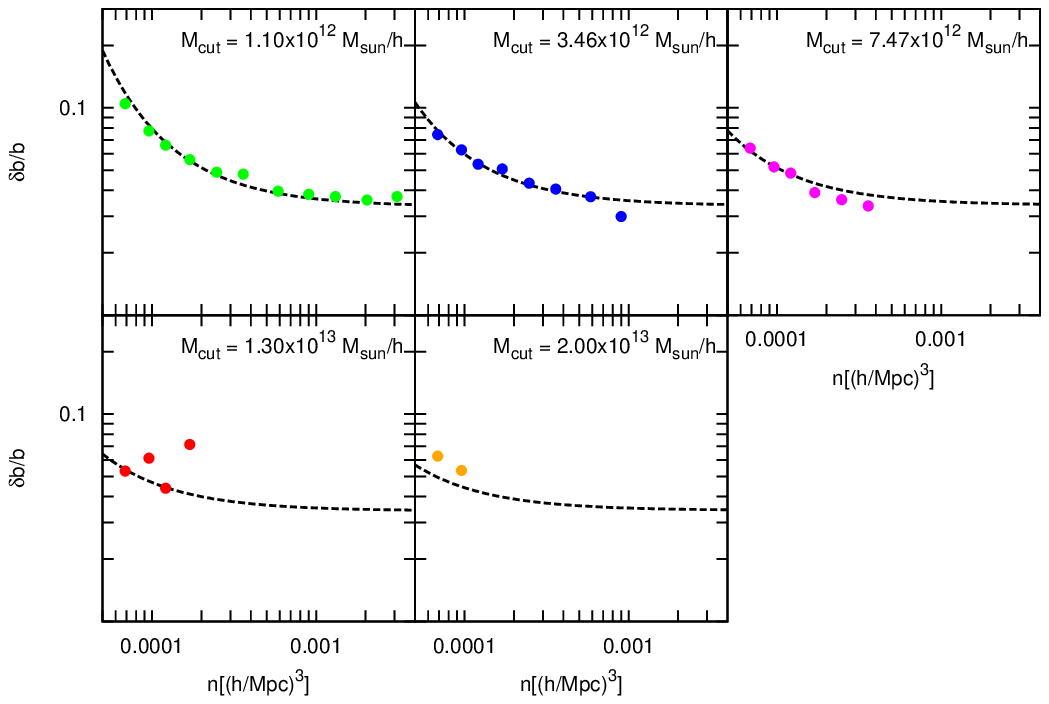}
  \end{center}
  \caption{Relative errors on bias $\delta b/b$ as a function of
    density $n$ for different mass (i.e. bias) samples, as labelled in
    the panels. The dots represent the mean error over the $27$ mock
    catalogues. The black dashed lines show the scaling formula of
    Eq.~(\ref{func}). The colour code is the same of the previous
    figures.}
  \label{bE_vs_density}
\end{figure*}
\begin{figure*}
  \begin{center}
     \includegraphics[width=16cm]{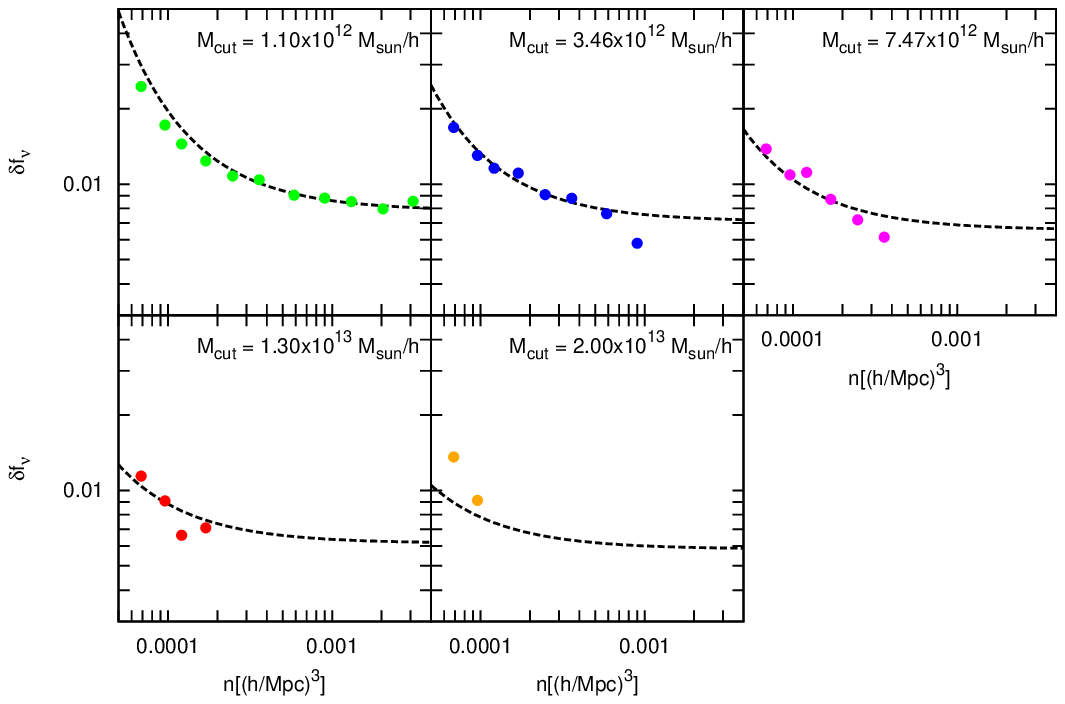}
  \end{center}
  \caption{As Fig.~\ref{bE_vs_density} but for errors on neutrino mass
    fraction, $\delta f_{\nu}$, as a function of density, $n$, for
    different mass (i.e. bias) samples.}
  \label{fE_vs_density}
\end{figure*}

In Fig.~\ref{bfnu_growth_comp} we compare the results obtained
considering $f(z) = \Omega_m^{\gamma}(z)$ (green contours) with the
ones obtained assuming $f(z) \simeq
[\Omega_{m}(z)(1-f_{\nu})]^{\gamma}$ on the scales of interest in this
work, {\emph i.e.} $k>>k_{FS}$ (as suggested by
\citealt{2008PhRvD..77f3005K}).  The results are very similar, indeed
averaging the errors obtained in the two cases, over the $27$ mock
catalogues, we find that the difference in the $f_{\nu}$ error is only
$\sim 12\%$.

As verified by previous works in the literature, the RSD model used
for this analysis is not sufficiently accurate at small non-linear
scales, especially for what concerns the quadrupole moment (see
e.g. \citealt{2015arXiv150501170M}).  Therefore, in order to minimise
systematic errors due to theoretical uncertainties, we consider only
scales larger than $ 7.5 \ \mathrm{Mpc}/h$, though our final results
are not significantly affected by this choice, considering the
estimated uncertainties.

\subsection{Error dependence on the survey parameters}
Having analysed all the samples in Table \ref{tab sub-samples}, we
can now present the results on the dependence of the errors on
three different parameters characterising a survey: bias, density and
volume.  First, we illustrate the dependence on one single parameter
at a time, and then combine these dependencies to provide a fitting
formula able to describe the overall behaviour.

\subsubsection{Error dependence on bias}
In Figs.~\ref{bE_vs_bias} and \ref{fE_vs_bias}, we plot the relative
errors on $b$ and $f_{\nu}$, respectively, as a function of bias, in
different density bins.  For all the samples considered, the volume is
taken fixed.  The error dependence on the bias is approximately
constant in the density range $1.7\times10^{-4} \ (h/\mathrm{Mpc})^3 <
n < 3.1\times10^{-3} \ (h/\mathrm{Mpc})^3$. For densities smaller than
$1.7\times10^{-4} \ (h/\mathrm{Mpc})^3$, the error decreases
as the bias increases.  In the high-density regime, the trend of the
error can be described by a power law of the form:
\begin{equation}
    \delta x \propto b^{\alpha_1} \ .
    \label{eq_b_ps} 
\end{equation}
In the low-density regime, that is below $1.7\times10^{-4}
\ (h/\mathrm{Mpc})^3$, the dependence is better described by an
exponential decrease:
\begin{equation}
    \delta x \propto \exp(1/b^{\alpha_2}) \ .
    \label{eq_b_exp}
\end{equation}

These results can be explained as follows.  At high densities the
errors on $b$ and $f_{\nu}$ are similar for all values of $b$.  At low
densities, the gain due to a high distortion signal of the low-bias
samples is cancelled out by the dilution of the catalogues. Instead,
the high-bias samples, which are characterised by a stronger
clustering signal and are intrinsically less dense, give a smaller
error and then are more suitable when estimating these parameters
using the correlation function both in the real and redshift space.

\subsubsection{Error dependence on density}
The dependence of the errors on the survey density is shown in
Figs.~\ref{bE_vs_density} and \ref{fE_vs_density}, for $b$ and
$f_{\nu}$, respectively.  We plot the errors estimated with samples of
different bias and density, having fixed the volume.  Both the errors
decrease exponentially, becoming constant for high values of the
density.  Indeed, a decrease in the density leads to larger errors,
due to the increasing shot noise, whereas moving to higher
measurements tend to become cosmic-variance dominated and the errors
remain almost constant. This behaviour can be described by an
exponential function of the form:
\begin{equation}
    \delta x \propto \exp(n_0/n) \ ,
    \label{eq_den}
\end{equation}
where $n_0$ is the density value that separates the shot noise regime
from the cosmic variance one.  We can notice that this exponential
decrease depends also on bias, with a flattening of the exponential
function for high-bias samples, reflecting what already observed in
the previous section.  Therefore, it is more appropriate to describe
these errors with a function that is a combination of
Eq.~(\ref{eq_b_exp}) and Eq.~(\ref{eq_den}):
\begin{equation}
    \delta x \propto \exp[n_0/(n b^{\alpha_2})] \ .
    \label{eq_b_den}
\end{equation}

\begin{figure*}
    \begin{center}
        \includegraphics[width=16cm]{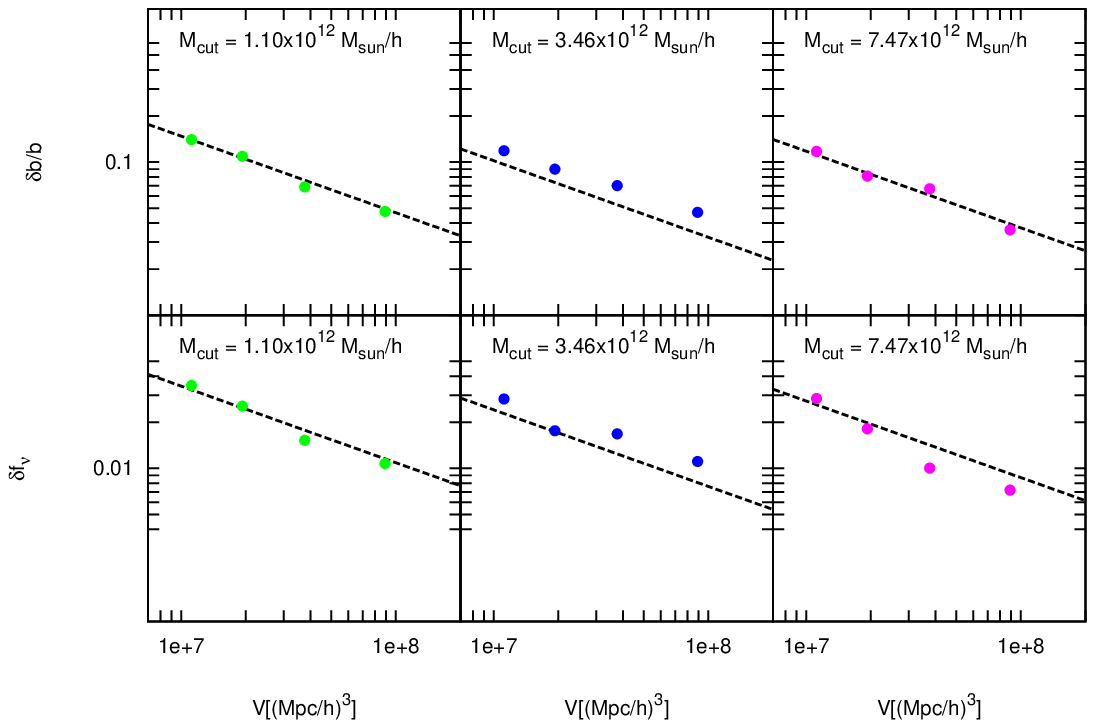}
    \end{center}
    \caption{Relative errors on the halo bias, $\delta b/b$ (left
      panel) and errors on neutrino mass fraction, $\delta f_{\nu}$
      (right panel) as a function of volume, for different mass
      samples (all labelled in the panels). As in the previous figures,
      the dots represent our measurements and the dashed lines show
      the fitting formula of Eq.~(\ref{func}).}
    \label{ErrVsVol}
\end{figure*}

\subsubsection{Error dependence on volume}
Finally, we illustrate the dependence on volume. We consider $5$
sub-samples of different bias and density, and for each of them we
split the cube of the simulation in $N^3$ cubes with $N=\{4,5,6\}$, in
order to reduce the volume of the catalogues.  We apply the same
method described before and compute the mean errors for each
sub-sample.  We find that the errors scale as the inverse of the
square root of the volume, irrespective of bias and density, obtaining
for $b$ and $f_\nu$ the same dependence found by
\cite{2008Natur.451..541G} and \cite{2012MNRAS.427.2420B} for $\beta$.
The results are shown in Fig.~\ref{ErrVsVol}, where we plot the
measurements from catalogues with different volume and bias values,
for a fixed number density.

\subsection{Fitting formula for the overall error dependence}
According to these considerations, we try to fit the errors with the
same functional form proposed by \citet{2012MNRAS.427.2420B} to
describe the error on $\beta$:
\begin{equation}
    \delta x \approx C b^{\alpha_1} V^{-0.5} \exp \bigg(\dfrac{n_0}{b^{\alpha_2} n}\bigg) \; .
    \label{func}
\end{equation}
We find that Eq.~(\ref{func}) can describe accurately also the errors
on $f_{\nu}$ and $b$.  The dashed lines in
Figs.~\ref{bE_vs_bias}-\ref{ErrVsVol}, 
represent surfaces of Eq.~(\ref{func}).  In
particular, in Figs.~\ref{bE_vs_bias} and \ref{fE_vs_bias}, the
dashed lines show Eq.~(\ref{func}) for fixed values of volume $V =
8.9\times 10^7 \ (h^{-1} \ \text{Mpc})^{3}$ and density $n$ (according
to the labels of each panel).  In Figs.~\ref{bE_vs_density} and
\ref{fE_vs_density}, the volume $V = 8.9\times 10^7 \ (h^{-1}
\ \text{Mpc})^{3}$ and the bias $b$ are kept fixed.  Finally, the
lines in Fig.~\ref{ErrVsVol} show the errors given by
Eq.~(\ref{func}).

The obtained best-fit parameters for Eq.~(\ref{func}) are: $C= 311
\ h^{-1.5}$ $\text{Mpc}^{1.5}$, $\alpha_1 = 0.1 $ and $\alpha_2 = 1.9$
and $C= 72 \ h^{-1.5} \ \text{Mpc}^{1.5}$, $\alpha_1 = 0.2$ and
$\alpha_2 = 2$ for the errors on $b$ and on $f_{\nu}$, respectively.
In both cases we assume $n_0=1.7 \times 10^{-4} \ h^3
\ \text{Mpc}^{-3}$, which is roughly the density at which cosmic
variance starts to dominate.  The errors that we fit are the relative
error for $b$, and the absolute error on $f_{\nu}$. Therefore, in the
fitting formula of Eq.~(\ref{func}), $\delta x$ should be replaced
with $\delta b/ b$ and $\delta f_{\nu}$, respectively.

\begin{figure*}
    \begin{center}
        \subfigure{\includegraphics[width=8.5cm]{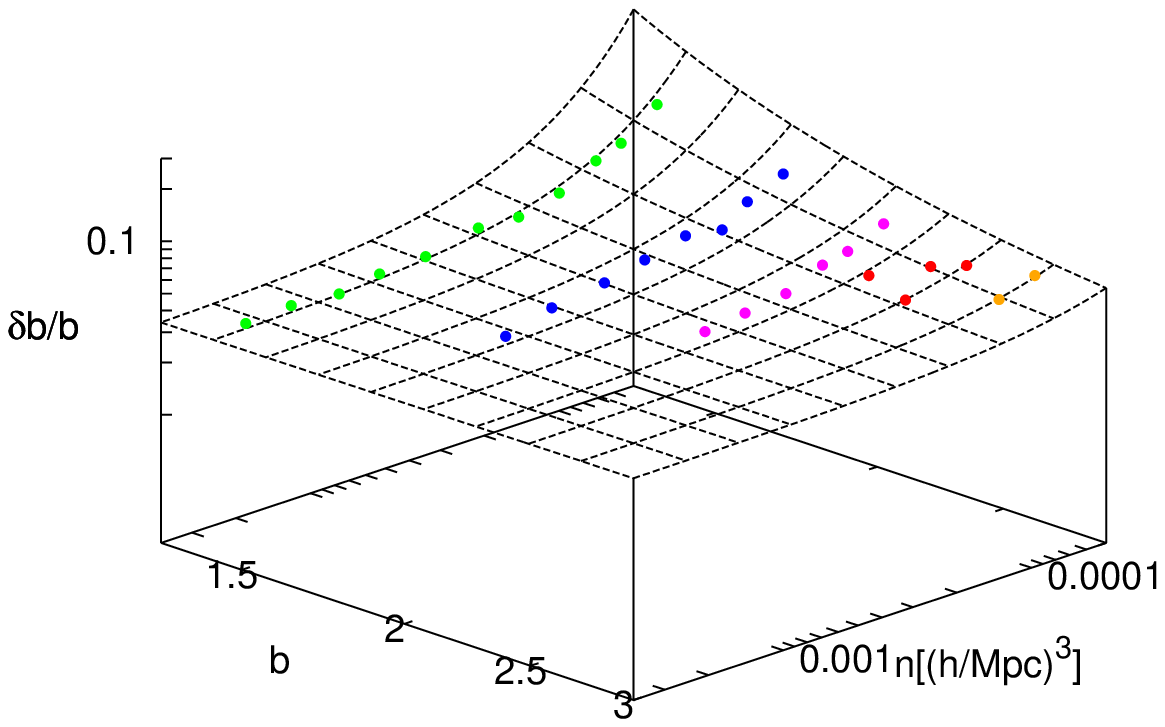}}
        \subfigure{\includegraphics[width=8.5cm]{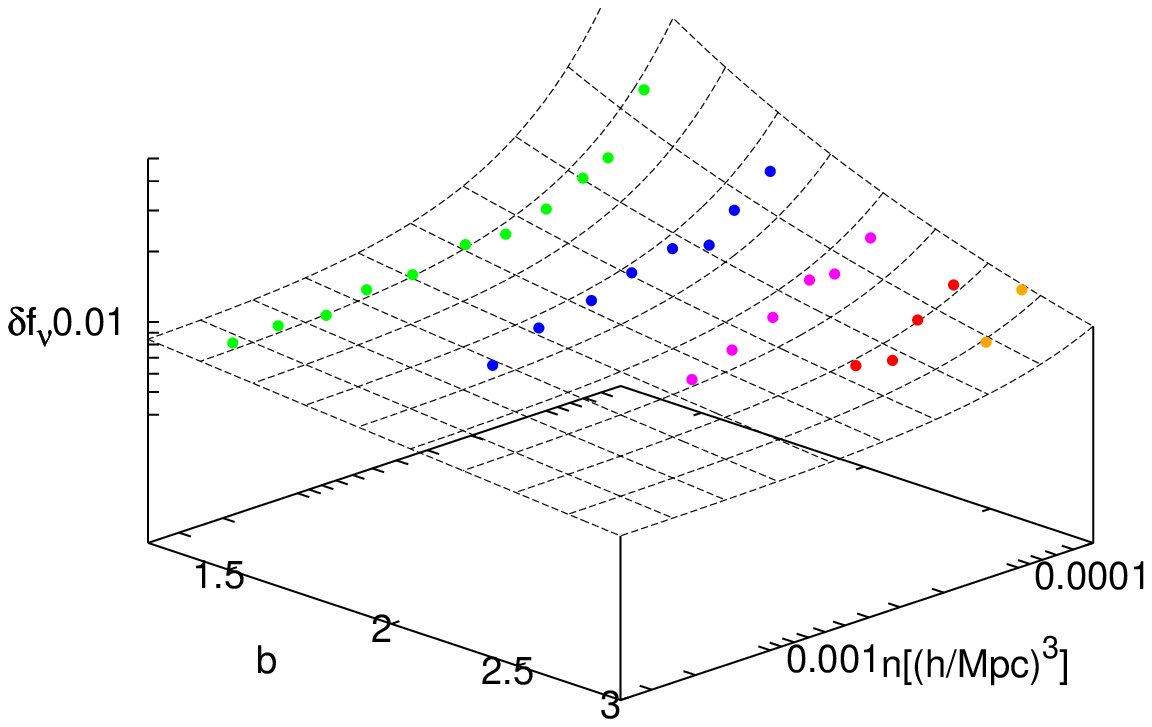}} \\
        \subfigure{\includegraphics[width=8.5cm]{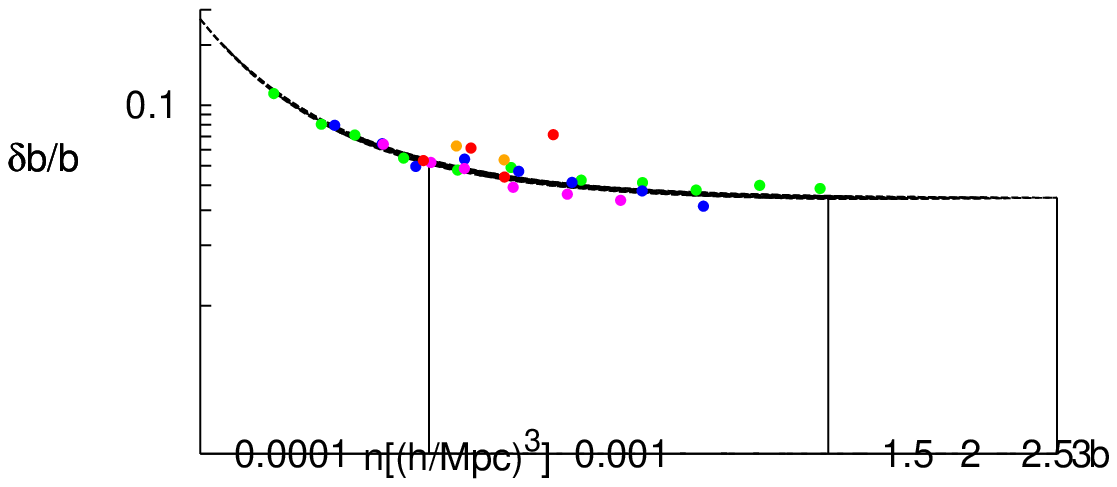}}
        \subfigure{\includegraphics[width=8.5cm]{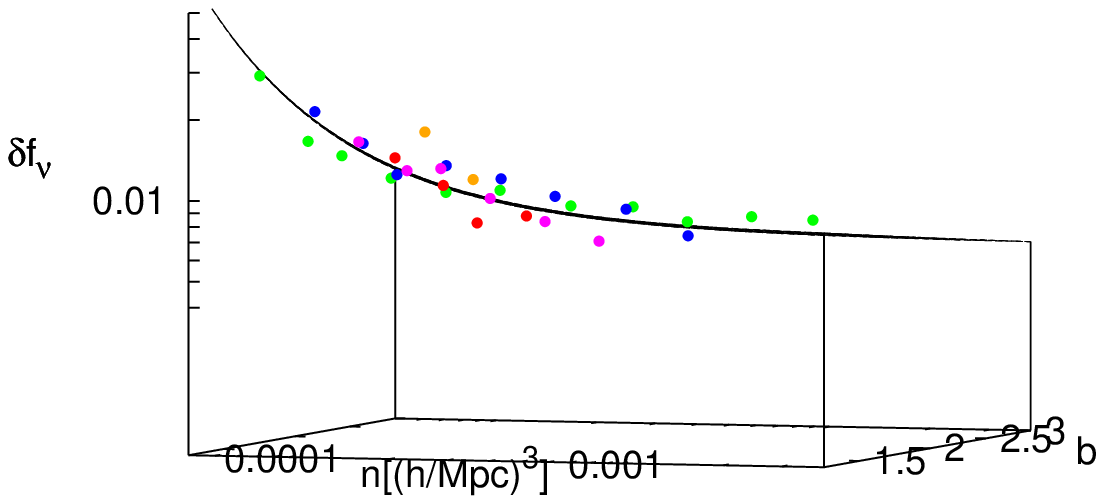}}
    \end{center}
    \caption{Top panels: relative errors on the halo bias, $\delta
      b/b$ (left) and errors on neutrino mass fraction, $\delta
      f_{\nu}$ (right), as a function of density, $n$, and bias, $b$,
      over-plotted on the surface described by the fitting formula of
      Eq.~(\ref{func}) for a fixed volume $V = 8.9\times 10^7 \ (h^{-1}
      \ \text{Mpc})^{3}$. Bottom panels: the same as the ones in top
      panels, except for the fact that the axes are oriented in order
      to highlight the agreement between our measurements and the
      fitting function. The colour code is the same of the previous
      figures.}
    \label{bfE_vs_b_and_d}
\end{figure*}

The overall behaviour of both errors is summarised in
Fig.~\ref{bfE_vs_b_and_d}.  In the top panels we plot the error on $b$
and $f_{\nu}$ as a function of density and bias for a fixed volume.
The dashed surface represents the fitting function of Eq.~(\ref{func})
with $V = 8.9\times 10^7 \ (h^{-1} \ \text{Mpc})^{3}$.  The bottom
panels show the same points, but suitably oriented to highlight the
agreement with the fitting function of Eq.~(\ref{func}).

\section{Summary and Discussion} \label{sec concl}
We have performed an extended analysis to forecast the statistical
errors of the neutrino mass fraction and the bias parameter exploiting
the correlation function in the redshift space.  We have measured the
multipoles of the correlation function in bins of $5 \ \mathrm{Mpc}
\ h^{-1}$, up to a scale of $35 \ \mathrm{Mpc} \ h^{-1}$, from mock
data extracted from the halo catalogues of the {\small BASICC}
simulation at $z=1$.  The halo catalogues have been selected in order
to have different values of bias, density and volume, that are three
fundamental parameters used to describe a redshift survey.

The mock data have been analysed using an MCMC likelihood method with
$f_{\nu}$, $b$ and $\sigma_{12}$ as free parameters, fixing all other
parameters to the input value of the simulation. We have presented the
results concerning only $f_{\nu}$ and $b$, considering $\sigma_{12}$
just as a nuisance parameter needed to take into account the effect of
non-linear motions.  The best-fit values for these two parameters are
in agreement with the input values of the simulation within $1\sigma$
for each considered sample.

The scale-dependent suppression in the power spectrum induced by
massive neutrinos would allow to constrain separately $f_{\nu}$ and
$b$. However, this effect is quite small, and it is difficult to
extract these constraints from the real-space clustering alone, due
to current measurement uncertainties. On the other hand, they can
be efficiently extracted from the redshift-space monopole of the
correlation function, as shown in Fig.~\ref{multi02}. Indeed, as
explained in \S \ref{MCMC}, while $f_{\nu}$ enters the model only
through the shape of the real-space undistorted correlation
function, the bias enters the model twice, both in the real-space
correlation function of matter and in the multipole expansion
through $\beta$. The quadrupole multipole has larger errors with
respect to the monopole. Still it can be exploited to improve our
measurements as the constrain direction is slightly different (see
Fig.~\ref{multi02} and Eqs.~\ref{eq mult}). Thus, as we have shown,
the use of both monopole and quadrupole together can help in
breaking the degeneracy between the halo bias and $f_\nu$.

For what concerns the error trend as a function of density, volume
and bias, we found that our measurements are fitted to a good
approximation by the scaling formula given in Eq.~(\ref{func}) for
both $\delta b/b$ and $f_{\nu}$.

A crucial point in this work is represented by the covariance matrix.
We have decided to use only its diagonal part. Though the
off-diagonal elements are not negligible, they are also very noisy due
to the small number of mock catalogues available, compared to the
number of bins used to compute the correlation function.  However, the
results presented in this work are not biased by the use of the
diagonal matrix.  As shown in the appendix, the full covariance matrix
introduces just a slight shift in the fitting function, and it does
not alter its form.

Some aspects still need to be investigated.  An improvement of the
fitting formula including a redshift dependence would be desirable.
Moreover, having a larger number of simulations with different
$\sigma_8$ can be useful to check if the variation of this parameter
could affect the error on $f_{\nu}$.  Finally, according to recent
works (\citealt{2014JCAP...02..049C,2015arXiv150705102V}), it would be
better to consider a linear bias defined as $b^2 \equiv
\xi_{halo}/\xi_{cdm}$.  However, when considering small neutrino
masses, the error caused by the assumption of a linear bias defined in
terms of $\xi_{m}$, instead of $\xi_{cdm}$, is negligible considering
the estimated errors of this analysis (see \cite{Castorina_etal2015}
for details about the effect of this choice on growth rate
estimations).

\begin{table}
  \caption{Forecast errors on neutrino mass fraction obtained with
    the fitting function given by Eq.~(\ref{func}) for some future
    galaxy surveys, assuming a bias factor $=1$.}
  \begin{center}
    \begin{tabular}{ccccc}
      \toprule
      $ $ & $V [(\text{Mpc}/h)^3]$ & $n \ [h^3 \ \text{Mpc}^{-3}]$ & $ \delta f_{\nu} $ \\
      \midrule
      $\text{Euclid}$ & $1.6\times10^{10}$ & $1\times10^{-3}$ & $6.7\times10^{-4}$\\
      $\text{WFIRST}$ & $1\times10^{10}$ & $1.3 \times 10^{-3}$ & $2\times10^{-4}$\\
      $\text{DESI}$ & $4\times10^{10}$ & $4.2 \times 10^{-4}$ & $4\times10^{-4}$\\
      \bottomrule
    \end{tabular}
  \end{center}
 \label{tab errors}
\end{table}

Regardless these still open issues, the presented fitting formula can
be used to forecast the precision reachable in measuring the neutrino
mass fraction with forthcoming redshift surveys.  Recent constraints
on neutrino mass came from different cosmological probes.  For
example, the latest Planck results \citep{2015arXiv150201589P} put an
upper limit on the sum of neutrino masses $\sum m_{\nu} <
0.23\ \mathrm{eV} $, and, in combination with LSS surveys, the
following constraints have been obtained: $\sum m_{\nu} <
0.18\ \mathrm{eV} $ \citep{2014PhRvD..89j3505R}, $\sum m_{\nu} <
0.29\ \mathrm{eV} $ \citep{2012JCAP...06..010X} and $\sum m_{\nu} =
0.35\pm0.10\ \mathrm{eV}$ \citep{2014MNRAS.444.3501B}.  If we consider
that a Euclid-like survey should be able to cover a volume of $V
\approx 1.6\times10^{10} (\mathrm{Mpc}/h)^3$, targeting a galaxy
sample with bias $b \approx 1.3$ and density $\approx 10^{-3}
(h/\mathrm{Mpc})^3$, the neutrino mass fraction can be measured with a
precision of $\approx 6.7\times10^{-4}$.  This value translates into
an accuracy of $\delta \left( \sum m_{\nu} \right) \approx
0.0081\ \mathrm{eV}$, smaller than the one quoted into the Euclid Red
Book \citep{2011arXiv1110.3193L}, obtained with the Fisher Matrix
method from BAO measurements.  This is mainly due to the fact that our
predictions have been derived using very different probes and
methodology, but most of all because we kept fixed many of the
relevant cosmological parameters such as, for example, $\Omega_m$ and
the initial scalar amplitude of the power spectrum \citep[see
  e.g.][]{Carbone_etal2011}.  Predictions for other surveys are
reported in Table \ref{tab errors}.  Overall, our analysis confirms
that the two-point correlation function in redshift space provides a
promising probe in the quest for neutrino mass.

\section*{Acknowledgements}
We acknowledge the financial contributions by grants ASI/INAF
I/023/12/0 and PRIN MIUR 2010-2011 ``The dark Universe and the cosmic
evolution of baryons: from current surveys to Euclid''.  The
computations of this work have been performed thanks to the Italian
SuperComputing Resource Allocation (ISCRA) of the Consorzio
Interuniversitario del Nord Est per il Calcolo Automatico (CINECA).
FP warmly thanks Mauro Roncarelli for useful
discussions. C.C. acknowledges financial support from the INAF
Fellowships Programme 2010 and the European Research Council through
the Darklight Advanced Research Grant (n. 291521).

\appendix
\section{Assessing the validity of the covariance matrix} \label{app}
The results presented in this work have been obtained considering only
the diagonal elements of the covariance matrices. Here we briefly
review the reasons that brought us to this choice. In order to test
the effects introduced by different covariance matrix assumptions, we
repeat our analysis using three different matrices, the diagonal
matrix, the full matrix and the smoothed matrix, the last one obtained
with a smoothing algorithm that follows the approach presented in
\citet{2012MNRAS.426..226C}. Specifically, the latter algorithm
exploits the fact that the diagonal elements of the covariance matrix
are larger than the first off-diagonal elements, than in turn are
larger than all other elements (see Fig.~\ref{covmat_02}).
Therefore, we consider the vector made up by the diagonal elements
only and average each of them using the two nearby elements, according
to the formula:
\begin{multline}
    \tilde{C}(i,i)  = (1-p)C(i,i) + \\
                        p \left[ C(i+1,i+1)+C(i-1,i-1) \right] /2 ,
\end{multline}
where $p$ is a weight. If one of the two nearby elements is not
present (i.e. when we consider the first and the last element of the
vector), then $\tilde{C}(i,i) = C(i,i)$.  The same algorithm is
applied to the first off-diagonal elements, while the ``generic"
elements of the covariance matrix are averaged using all the nearby
elements:
\begin{multline}
    \tilde{C}(i,j) = (1-p)C(i,j) + \\
            \frac{p}{m} \cdot
            \begin{bmatrix} 
            C(i+1,j)+C(i-1,j) + \\
            C(i,j+1)+C(i,j-1) + \\
            C(i+1,j+1)+C(i-1,j-1) + \\
            C(i+1,j-1)+C(i+1,j-1)
            \end{bmatrix} ,
\end{multline}
where $m$ is the number of nearby elements used in the averaging
procedure.  For all the matrix elements we used $p=0.01$. As verified,
this smoothing procedure helps to alleviate some of the numerical
problems related to the matrix noise, though it does not work properly
for all cases considered.

\begin{figure}
    \begin{center}
        \subfigure{\includegraphics[width=8.5cm]{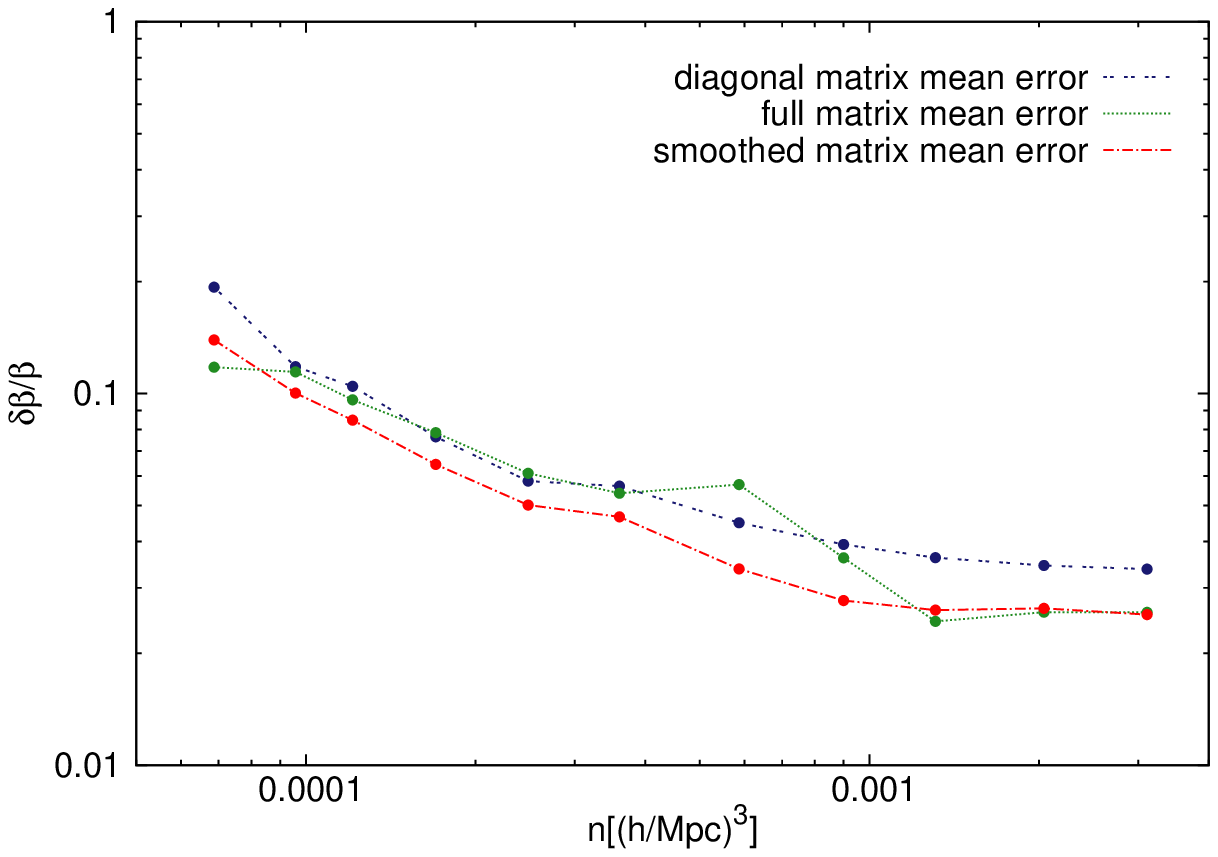}} \\
        \subfigure{\includegraphics[width=8.5cm]{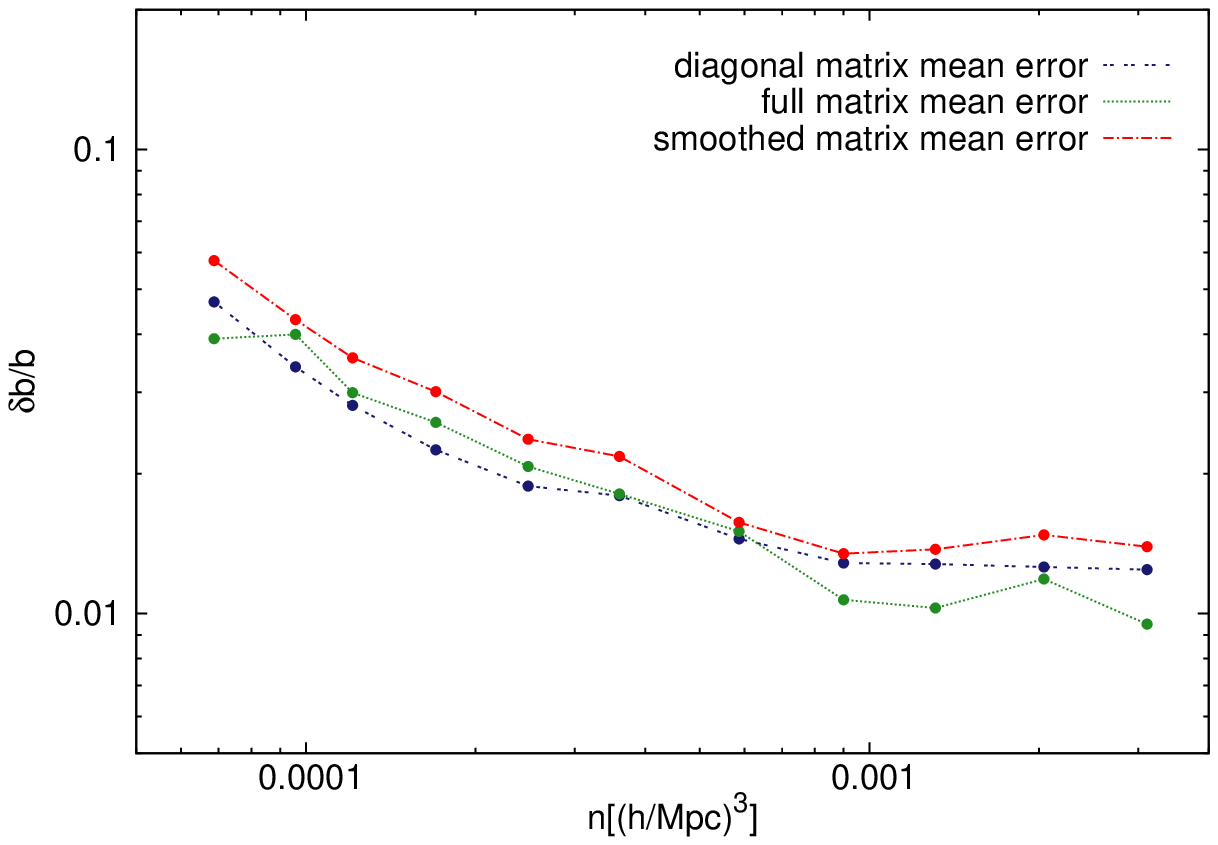}}
    \end{center}
    \caption{Relative error on the distortion parameter, $\delta
      \beta/\beta$ (upper panel), and bias, $\delta b/b$ (lower panel),
      as a function of density, $n$, obtained analysing the mock data
      with $\beta$ and $b$ as free parameters. The dots represent the
      MCMC error averaged over the $27$ mock catalogues extracted from
      the most dense sample with $M_{cut} = 1.1 \times 10^{12}
      \ h^{-1} \ M_{\odot}$, obtained using the diagonal matrix (blue
      dashed lines), the full matrix (green dotted lines) and the
      smoothed matrix (red dot-dashed lines).}
    \label{bb_err_vs_d}
\end{figure}
For these tests, we consider the simple case where the only free
parameters of the MCMC analysis are the distortion parameter, $\beta$,
and the bias, $b$. We choose this limited parameter space in order to
speed up the computation. Fig.~\ref{bb_err_vs_d} shows the errors on
$\beta$ and $b$ as a function of density, obtained with the diagonal
matrix (blue dashed lines), the full matrix (green dotted lines) and
the smoothed matrix (magenta dot-dashed lines). As it can be noted,
the shape of the curves is quite similar, while the normalisation is
slightly different. For instance, the differences between the errors
obtained with the diagonal and the full covariance matrix are $\sim
5\%$ for $\beta$ and $\sim 2\%$ for $b$.  However, the small number
of mock catalogues available to construct the covariance matrices,
relative to the number of bins analysed, does not allow us to get
robust results \citet{2007A&A...464..399H}. These reasonings, together
with the fact that using the diagonal matrix we get a less scattered
trend for the errors in all the cases considered, lead us to neglect
the non-diagonal elements of the covariances.

\bsp

\label{lastpage}

\end{document}